# In the Face (book) of Social Learning

**Michail N. Giannakos**
*Old Dominion University, Department of Computer Science, Norfolk, VA, USA,*
*mgiannak@cs.odu.edu*
**Patrick Mikalef**
*Ionian University, Department of Informatics, Corfu, Greece, mikalef@ionio.gr*

## ABSTRACT

Social networks have risen to prominence over the last years as the predominant form of electronic interaction between individuals. In an attempt to harness the power of the large user base which they have managed to attract, this study proposes an e-learning prototype which integrates concepts of the social and semantic web. A selected set of services are deployed which have been scientifically proven to positively impact the learning process of users via electronic means. The integrability of these services into a social network platform application is visualized through an exploratory prototype. The Graphical User Interface (GUI) which is developed to implement these key features is in alignment with User-Centered principles. The designed prototype proves that a number of services can be integrated in a user-friendly application and can potentially serve to gain feedback regarding additional aspects that should be included

## 1. INTRODUCTION

Learning related theories have dominated the interest of psychologists and academics for the past few decades. In literature, a vast amount of ambiguous opinions about how knowledge is optimally transmitted exist; all of which are based on unique traits. One of the pioneers in this field, Albert Bandura, proposed the idea of Social Learning (Bandura, 1986), a form of learning which places emphasis on the vicarious part of learning. According to his theory, knowledge is transmitted through the process of socializing, i.e. interacting with others and observing their behavior. However, according to scholars (Siemens, 2005), the efficacy of social learning is severely impeded by the amount of available information within the learning community, which is one of the major restraints. In terms of information enrichment this can be translated as the number of active participants in a social learning community.

This characteristic of social learning in conjunction with the continuously growing social network user base could leverage the potential of a next generation of e-learning. The significance of social networks for e-learning purposes has been a subject of quite some attention recently by scholars (Cuéllar et. al., 2010; English & Duncan-Howell, 2008, Mazer et. al., 2007). Results from empirical studies conducted on the effectiveness of social networks as a learning medium, have concluded that not only do they produce positive learning outcomes but also promote the social interaction between the tutor and the learner. In recent years, users' intention to adopt such social networks has increased incrementally with some prominent examples being Facebook, Twitter and MySpace (Mikalef et. al., 2012). This fact provides a stepping stone to wok upon.

However, social networks are not the only evolution of the World Wide Web (WWW) which could potentially contribute towards learning. In contrast with traditional web technologies (*Web 1.0*) that provided an asynchronous method of sharing data over the WWW, semantic web (a component of *Web 3.0*) allows for the creation of intelligent services that perform tasks on behalf of users. The main idea behind the semantic web concept is to treat the WWW as a pool of data from which data can be extracted and combined in order to create new knowledge. It is evident that technologies based on semantic web concepts, provide a powerful tool to extract structured information from the web without having to deal with information overload.

Despite the rapid growth of both social networks and semantic web technologies, little attention has been set on the combined potential of these two technologies in contributing to education and more specifically to the learning process. Previous efforts have focused primarily on the potential of social networks to promote learning (Ozkan & McKenzie, 2008; English & Duncan-Howell, 2008; Conole & Culver, 2009) or on the effect that interactive information sharing technologies (Semantic Web) have in reshaping existing knowledge distribution methods (Gladun et. al., 2009; Dutta, 2006; Henze et. al., 2004). The use of social learning networks to promote e-learning has been subject to limited attention so far. Despite this, the few empirical studies conducted which have centered on this field; conclude that the contribution of these technologies is beneficial for all participating parties (English & Duncan-Howell, 2008; Liccardi et. al., 2007). This positive relationship in also confirmed when deploying semantic web technologies in order to enhance learning. In a study by Koper it is found that semantic web technologies can significantly promote learner-centered education for lifelong learning without increasing the workload for learners and staff (Koper, 2004).

In this paper we aim at examining how the wide adoption of social networks could serve as a means of promoting collaborative and lifelong learning by incorporating a number of semantic web features. By combining existing scientific literature on the two fore mentioned domains, we aim at answering the following question:

*"How can state of the art web technologies be integrated in such a manner in order to promote lifelong social e-learning?"*

Given this research gap that exists in the combination of the two technologies, we propose a design of an exploratory prototype, focused on an integrated e-learning social network (e-LSN). This prototype aggregates the large user base of a social network and the functionalities of emerging semantic web technologies. The proposed prototype could be deployed on any social network - in our illustrated prototype we chose Facebook due to its vast adoption by users - with a number of semantic web and multimedia features in order to make it more tangible. These include: social content management (synchronous/asynchronous), web-cast media, wiki-enabled information handling and intelligent web and file search. The aim of this paper is twofold; a) to overview and analyze the potential of social networks and semantic web technologies in learning b) to present an exploratory prototype, depicting the feasability of aggregating these two technologies, in order to gain feedback and aspire new ideas.

## 2. BACKGROUND AND RELATED WORK
### 2.1. Web evolution in education services

The World-Wide Web (WWW) is an important learning technology platform. Its accessibility has made it a successful environment in particular for educational purposes. Learning resources can be provided in several formats that can be accessed at any time and from any location. The WWW, however, is undergoing an evolutionary phase. This metamorphosis of the Web can potentially have a significant impact on the methods and applications of education.

*Web 1.0* is the term used to refer to the early stages of the WWW. *Web 1.0* consisted of services that were limited to data sharing over the internet and the representation of static websites (O'Reilly, 2005). As such, *Web 1.0* was a read-only medium; because of its nature, hence, *Web 1.0* could not satisfy the educational needs to the degree experienced in a conventional educational setting. This is the main reason why *Web 1.0* had limited educational applications.

The evolution of *Web 1.0*, i.e. *Web 2.0* has been defined according to the statement *"the Web of documents has morphed into a Web of data…we're looking to a new set of tools to aggregate and remix microcontent in new and useful ways" (*MacManus & Porter, 2005). This new generation of WWW is characterized by both the highly efficiency and interactivity among user and interface designs as well as the "ability to harness collective intelligence" (O'Reilly, 2005). For instance, web applications, such a wikis, blogs and forums illustrate the current trend toward the use of collaborative exploration and contribution of information and commentary in popular culture, as well as academic environments.

As such, *Web 2.0* can be simply defined as a read/write medium. This Transformation (from read only into read write medium) also affects the area of education. New areas of interest such as E-learning 2.0, Classroom 2.0 and Enterprise 2.0 are rapidly emerging paradigms (McAfee, 2006). However, several *Web 1.0* applications have been developed by identifying needs of user's in interacting and manipulating the educational content (Figure 1).

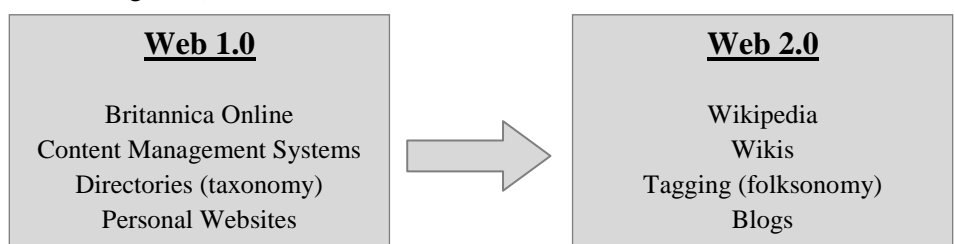

*Figure 1 The Impact of Web Evolution in Applications*

In recent years, education concepts based on *Web 3.0* - Semantic web services (Tie-Jun, 2009; Vega-Gorgojo, 2010) have appeared. *Web 3.0* and semantic web offers more intelligent services and in addition to reading and composing content, user's actions can initiate web processes, which can be possible with technologies such as, smart interfaces and intelligent-tutoring agents.

## 2.2.    Web 3.0 – Semantic web in education

Web-based systems are facing challenges that include: extensibility, interoperability, contextualization and consistency of metadata, integration and reuse of content, distribution of services, and so on. Such challenges are related to the attempt to represent the information on the Web in a way that computers can understand and manipulate it. Research in this direction is known as *Web 3.0* - Semantic Web (W3SW) research. Roughly speaking, W3SW extends the classical Web by providing it with a semantic-intelligent structure of web pages to give support to human and artificial agents to understand the content. As a result, the W2SW provides an environment where software agents can navigate through documents and execute sophisticated tasks. W3SW enables numerous improvements in the context of distance education systems contributing to the upgrade of learning quality. Nowadays, several researches (e.g., Bittencourt, 2010; Vega-Gorgojo, 2010) have used W3SW into educational approaches with great success.

## 2.3.    Social Networks in education

Several recent papers (e.g., Conole, 2009; English & Duncan-Howell, 2008; Lockyer, 2008) have examined the opportunities of social networks in developing learning environments. Among the most popular social networks are the facebook.com, twitter.com and myspace.com websites which enable users to provide social content in different ways. In many cases, the aforementioned social networks have successfully enhanced learning strategies (Joshua McCarthy 2010; English & Duncan-Howell, 2008).

In educational contexts, social networking behavior is related to learning by creating systems of information, contacts and support (Angela Yan Yu, 2010). Additionally, a great number of studies (Table

1) have demonstrated that social networking sites can valuable support students undertaking teaching practicum. Furthermore, it has been illustrated in literature that students with increased engagement in online social networking are more likely to have better affective development and learning success (Steinfield et al., 2008).

In this paper, we aim to integrate the use of social network with the W3SW capabilities. In order to achieve that, we promote the semantic-intelligent structure of W3SW through the well adopted environment of a successful social network. In our research we have identified a number of specific challenges:

- How do we achieve critical mass and sustainability around an evolving community of social network with a shared interest in learning?
- How can we take the advantage W3SW capabilities throw a social network?
- How social networking can be a vehicle for gaining information and knowledge?

*Table 1 Social Networks and Semantic Web Studies in e-Learning Framework*

| Technology | Title | Context | Implications | Source |
|---|---|---|---|---|
| **Social Networks** | Cloudworks: Social networking for learning design | This paper describes a new social networking site, Cloudworks, which aims to provide a mechanism for sharing, discussing and finding learning and teaching ideas and designs. | The site achieves critical mass and is self sustaining through end-user engagement and contributions. | Conole and Culver, 2009 |
| **Semantic Web** | Use of the semantic web to solve some basic problems in education: Increase flexible, distributed lifelong learning, decrease teachers' workload | Two application areas for use are discussed: a) software agents that support teachers in performing their tasks, and b) software agents that interpret the structure of distributed, self-organized, self-directed learning networks. | The resulting information is used by learners to help persons to perform their tasks in this context more effectively and efficiently. | Koper, 2004 |
| **Social Networks** | Facebook Goes to College: Using Social Networking Tools to Support Students Undertaking Teaching Practicum | This paper explores the use of social networking tools, such as Facebook, to support students undertaking teaching practicum | This paper suggests how the digital behaviours and habits of students enrolled in courses and may be used in developing supportive tools. | English & Duncan-Howell, 2008) |
| **Web 2.0** | The good, the bad and the wiki: Evaluating student generated content for collaborative learning | This paper explores the potential for wiki-type open architecture software to promote and support collaborative learning. | Collaboration, rather than competition, should be emphasised as a key aim of any wiki-based activity. | Wheeler et al., 2008 |
| **Social Network** | I 'll see you on "Facebook": The effects of computer-mediated teacher self-disclosure on student motivation, affective learning, and classroom climate | This study examined the effects of teacher self-disclosure via Facebook on anticipated college student motivation, affective learning, and classroom climate. | This study suggests that when a teacher self-discloses certain information, such as personal pictures, messages from friends and family, and opinions on certain topics, students may perceive similarities between themselves and the instructor. | Mazer et al., 2007 |
| **Semantic web** | Semantic search of tools for collaborative learning with the Ontoolsearch system | This paper introduces "Ontoolsearch", a new search system that can be employed by educators in order to find suitable tools for supporting collaborative learning. | Results showed that retrieval performance was significantly better with Ontoolsearch, despite educators' previous experience with keyword searches. | Vega-Gorgojo, 2010 |
| **Social Network** | Integrating social networking technologies in education: a case study of a formal learning environment | This paper presents a case study that examines the social networks' technology and experience in a formal education context. | Results concluded that this experience of using a social networking site in a formal education environment realized positive learning outcomes and experiences. | Lockyer, 2008 |

## 3. PROPOSED SERVICES

Following successful paradigms of technology enhanced learning services, we propose using a number of such services, targeted in fulfilling specific learning objectives. We base our selection on empirical findings throughout the academic literature. Only services which have been explicitly tested and results prove that they have a positive impact on the learning process are deployed. These services include technologies of the *Web 2.0* as well as the *Web 3.0* domain.

### 3.1. Chat Communication

Web-chat is one of the most widely used applications on the web since its inception. Chat conversations tend to share some features of oral language despite being conveyed through textural messages (Koch & Österreicher, 1994). As a consequence, information can be shared through such a medium, therefore, it can be regarded as an informal education tool of the web (Pfister & Mühlpfordt, 2005). The value of chat communication in the learning process has been empirically validated by a number of scholars (Pfister & Mühlpfordt, 2002; Jepson, 2005).

Educational benefits of Facebook have been found to include its ability to connect learners with each other into new networks of collaborative learning. Perhaps the most revealing, and according to Pew (2007), the most used feature is the wall. The wall is in essence an asynchronous chat facility in which text as well as multimedia related comments can be added and discussed. Through this service, users can exchange short text messages with their friends, similar to other micro-blogging websites such as Twitter. Despite the recognized benefits of chat in e-learning, it is argued that new tools supporting multimedia materials are highly demanded in e-Learning systems (Schreiber and Berge, 1998).

### 3.2. Webcasts

Webcasting has emerged as one of the premier push technologies for delivering video and audio content. Bell (2003) proposed web-casting as a learning tool, while Evans (2008) claims that students are more receptive to learning material provided in the form of a webcast than a traditional lecture or textbook. Additionally he argues that students perceive the use of webcasts as a more effective revision tool than textbooks, and more efficient than their own notes in helping them to learn. Video lecturing have proven to be beneficial means to e-learning as exemplified by a number of empirical studies (e.g. Giannakos and Vlamos, 2012). More specifically, Traphagan et. al., (2009) conclude that even with a reduced class attendance, students learning experience and performance are greatly enhanced.

### 3.3. Intelligent Search via Metadata

Metadata also known as Metacontent are defined as data that provide certain information regarding one or more other data resources. Such tags are often used in multimedia content (Videos, music, pictures) as well as on websites. Metadata are primarily used in order to assist the retrieval and to provide semantics to a large amount of information.

A pedagogical (smart) agent which uses metadata form a registry and can return lectures, relevant blogs, books and webcasts about the topic to the learner (Ohler, 2008). Smart agents assist learners in searching, documenting and archiving learning products and working collaboratively (Anderson & Whitelock, 2004). Semantic web ontologies will link the learner's needs and characteristics so that pedagogical agents can search for learning material based on the learners' need (Stojanovic, Staab, & Studer, 2001). We argue for a separation regarding the content that is returned when applied in our proposed system, between web and multimedia content.

### 3.4. Wiki-enabled Text using the Resource Description Framework (RDF)

Wikis are widely adopted website-based knowledge management systems which have gained mainstream popularity primarily in educational settings. The power of any wiki lies in the user contribution to its content development. A prominent such example is Wikipedia, which has managed in a period of less than 10 years to become the largest multi-lingual online encyclopedia world wide. It has become a very frequent phenomenon in a number of websites to enable hyperlinks to Wikipedia based on specific keywords. Through this way buzzwords or unknown terms can be explained by means of the online encyclopedia to users rather than analyzing them in detail repeatedly. Combining this function with the Web 3.0 set of specifications which provide semantics to data of websites, can automate the process of linking keywords with the appropriate websites that contain the information. By selecting a number of knowledge-related websites that employ the RDF specifications, both the linking of keywords-to-content and content-to-content is made easier, hence reducing the time spent on manual information retrieval. Despite semantic web being still in its infancy, there have been some studies on how they can be used for information linking (Willighagen et. al., 2011; Barker & Campbell, 2010).

## 4. PROPOSED SYSTEM

We have integrated the proposed services of the previous chapter through the deployment of an low-fidelity exploratory prototype. Prototyping is defined in literature as a well-designed phase in the production phase, where a model is produced in advance, exhibiting all the essential features of the final product (Heyer and Brereton, 2010). More specifically, exploratory prototyping is a means of prototyping in which the target is to clarify the goals of a proposed system. In our work, we weren't satisfied with dichotomy between probe and prototype, thus preferring the term 'exploratory prototype' as some form of middle way. A low-fidelity exploratory prototype therefore, is in essence a static non-computerized visualization of a system.

MUSEC is a mock-up application running on the social network Facebook. It has been designed in order to promote lifelong learning of music through a popular social network. By using MUSEC, users have the choice of selecting one or more courses on musical instruments according to their preference. Recognizing that Facebook has the largest user base amongst other social networks (approximately 600 million as of January 2011), MUSEC is deployed in order to harness the power of this population.

The graphical user interface of MUSEC`s main page is presented in Figure 1. Emphasis was placed on providing an interface which is on the one hand not overloaded with features and on the other hand provides the user with the appropriate level of functionality. Dynamic content modification is one of the core aspects of the application, meaning that once a user selects an option from the menu, it is automatically - i.e. without having to refresh the browser content - visible in his monitor. The pane is divided into three segments, each containing a different type of functionality. The main window allows for registration or deregistration from a specific music course. The right pane under the title News Feeds, provides information regarding registered courses as well as about friend interactions with the platform, thus enabling a first level of indirect interaction. Finally the Friends Online frame, depicts all of the users friends which are currently using MUSEC. It should be noted that friends are regarded as individuals which are connected through the Facebook platform.

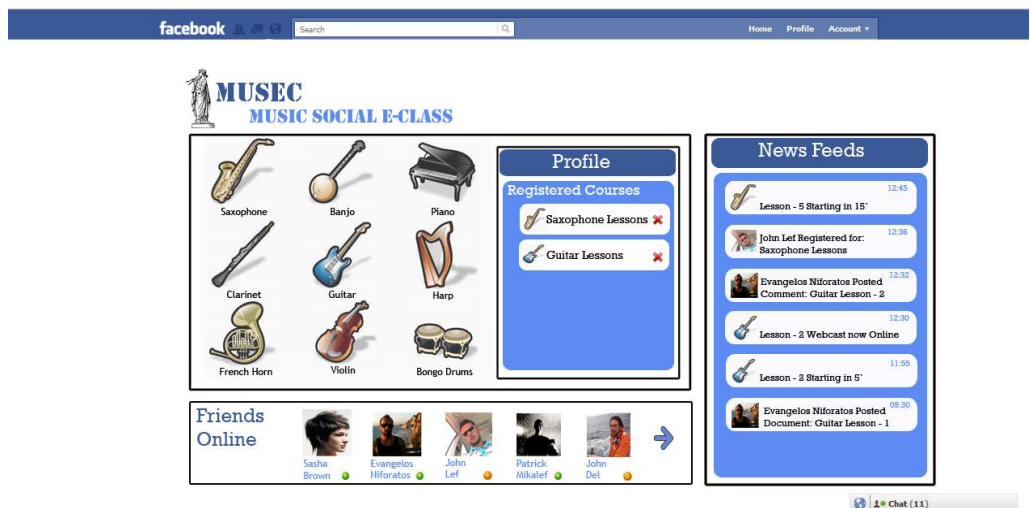

*Figure 2 Main page of MUSEC application*

One of the aspects that were taken into account in the design of the MUSEC platform, was that of feature overload. A recurring issue when operationalizing technology is that although a number of useful features may be applied, due to their abundance, their full functionality is not harnesed (Karr-Wisniewski and Lu, 2010). For this reason the complete list of features are apportioned in the three levels of MUSEC. The second GUI which users interact with is presented after a specific instrument course has been selected. Figure 2 depicts such an interface for the guitar course. In contrast with the previous view, the second level focuses primarily on a specific course, therefore the content is represented accordingly. The main window of Figure 2 illustrates the different levels of courses, divided into three main categories. For each of these categories, the lessons that are required to be attended are listed with a round colored icon indicating the state at which this course is. The application is built on the premise of supporting both synchronous and asynchronous e-learning, therefore, the different colorizations of the icons represent the state of the specific lesson (Blue = Asynchronous, Green = Synchronous, Red = No content available yet). In addition the Friends Online pane has been narrowed down to the individuals who are also following the specific course.

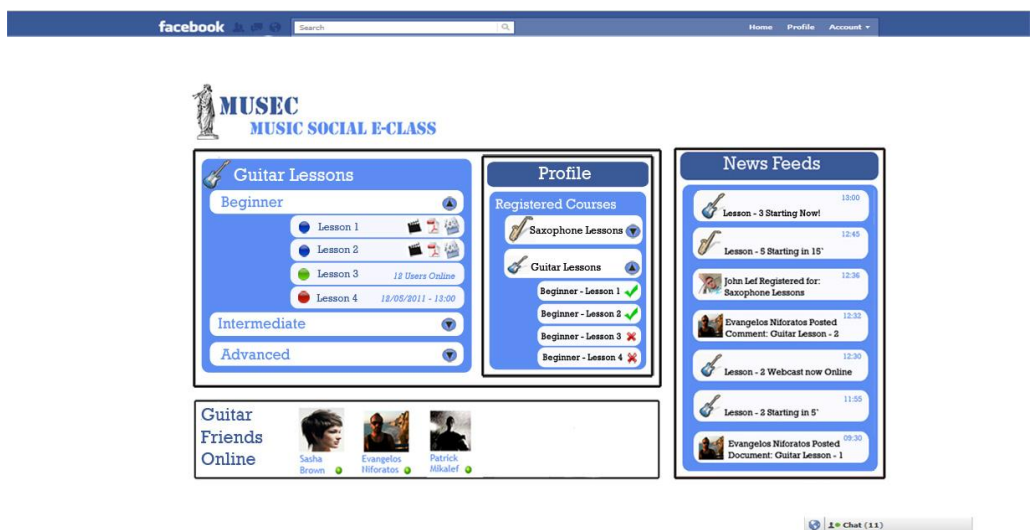

*Figure 3 Course level GUI of MUSEC*

In the course level GUI, users can track what lessons they have followed and select the ones which they have still to attend. Once selecting a specific lesson they are redirected to the third level (Lessons level) and final level of the MUSEC interface which incorporated all the necessary features to provide an interactive e-learning environment. Figure 3 as depicted below, presents an example of a synchronous lesson in which the tutor presents through a real-time web-cast, a guitar course. After the completion of the lesson the video of the web-cast is used for students which were not able to participate in the real-time lesson (Asynchronous). The text-box directly below the web-cast screen provides a description of the course and automatically links keywords to during either an ongoing or a past lesson, users are able to interact by means of a chat located in the leftmost side of the pane. Additionally, the option of directly posting a question to the tutor has been included, thus increasing the interactivity between tutor and learner. A key feature is the customized background live search, which enables students participating in a lesson, to explore a number of - related to the course - websites. These websites are returned both on the keywords entered by the user and on the metadata extracted from the web-page at hand content, thus providing more specific results. Finally, in the right pane users can download or add content which they believe is related to the course at hand. Federated search isapplied through the web crawl button, which return documents from a number of selected sources using existing document meta elements.

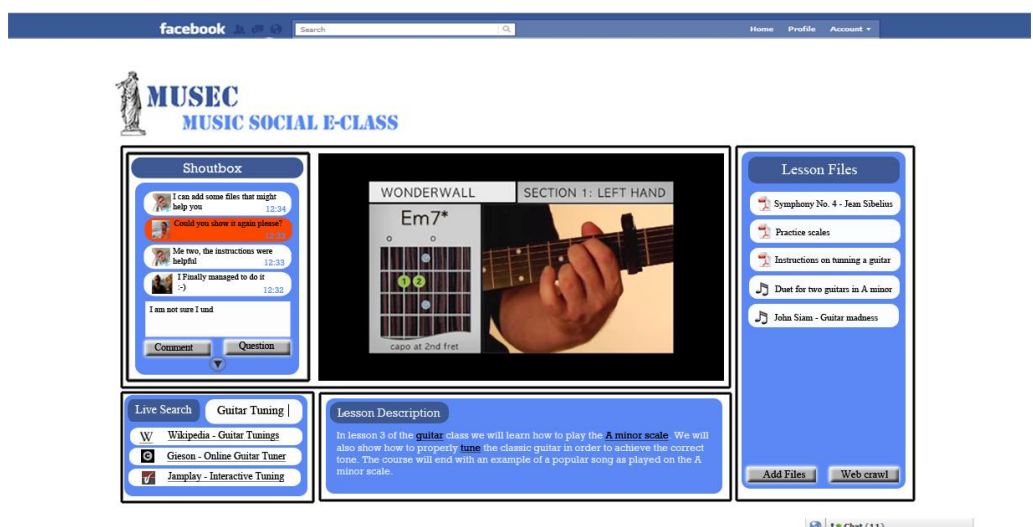

*Figure 4 Lesson level of MUSEC*

We conclude that the integration of the services we propose in order to enhance e-learning through a social network platform is feasible. However the proposed system still remains at a non-functional state, therefore, no empirical research can be performed in order to examine its impact on the learning process.

## 5. DISCUSSION

Educators seeking to introduce learners to social networking sites. However, many of them already have experience with Facebook and other social networking sites, several researchers indicate that learners' are already benefiting from the use of this engaging tool (e.g., English & Duncan-Howell 2008; Conole and Culver, 2009). For instance, the functionality to create groups on Facebook can be used for interest groups, study groups and task-team groups. Groups can keep learners' connected in a space in which they already feel comfortable, and allow them to post data relevant to the group, thereby facilitating learning. As most of the social network sites are opened to the public, any number of applications supporting e-learning and education can be implemented, opening the way for the tool to become increasingly useful from a pedagogical perspective.

Students benefit from the capabilities of W3SW. For example, semantic web search will return a multimedia report rather than just a list of hits, a smart agent can return webcasts, relevant blogs and books about the topic to the potential learner and ontologies will link learners' needs and characteristics and personalized agents can search for learning material based on the learners' needs.

The potential learners' may tend to be more interested in the social than the teaching dimensions of tools such as MUSEC. The technology undoubtedly helped to emphasise tutor's availability and helped learners' feel that they are part of a group and sharing the same learning environment. This higher degree of social presence may well be one of the greatest contributions of such tools.

MUSEC application provides both individual and collective benefits. This immediacy between action and positive outcome may very well create a positive reinforcement effect for the potential tutors' and learners'. First, any published result is visible and therefore potentially beneficial to others right away. As others see useful contributions being made, they can use these contributions, as well as build upon them and add their own associated knowledge. Jay Cross states in his book Informal Learning: Rediscovering the Natural Pathways that Inspire Innovation and Performance "workers learn more in the coffee room than in the classroom". Perhaps using this metaphor we can suggest that MUSEC can serve as a new coffee room.

It remains to be seen how useful the contributions were to the eventual product, measuring this contribution in each several dimensions is a challenge for the design team but the usefulness of the engagement, even at the base level of understanding the user group and the its basic interactions with the application is invaluable.

## 6. FUTURE WORK AND CONCLUSIONS

The Educational W3SW is still in its infancy, and instructors need to understand the potential of the emerging W3SW and possible implications for e-learning to prepare for the future when the Semantic Web is fully integrated into e-learning. In this paper, we describe a W3SW-based exploratory prototype approach aiming to introduce potential learners' into social and lifelong learning. The frequent usage of the application comes from its social functionalities and the adapted content which raises the interest.

In this paper, we have presented the design of a graphical user interface, MUSEC, which can serve as a basis for further development of social networking based W3SW applications. MUSEC provides an innovative experiential approach that supports social and lifelong learning. Our work extends prior research that explores the potential benefits of social networks or semantic web services in education in a general framework and forms up an exploratory prototype for the educational audience.

Furthermore, we showed that semantic web yields potential for research on technology-supported learning. By exploiting existing services and thanks to their integration, we assemble social networks and W3SW services into a prototype that allows us to make a first step towards the examination of its perspectives.

To summarize, our proposed system offers an intriguing and unparalleled wealth of functionalities. The exploitation of this functionality offers a high potential for the future of technology-enhanced learning. However, it is important not to forget that even if technology can be inspiring, the main focus in e-learning should still lie on the needs of the learner.

While we have designed a novel approach for integrating the capabilities of semantic, web and social networking, the questions of how effective the approach is and how usable the tool is for the task still remain. The next logical step would be a usability testing, to try to validate the approach with potential highly experienced Facebook users (experts). Once we have further refined the design of the prototype expect to apply it in vertical prototyping iteration. This will give us an opportunity to evaluate whether or not our prototype is effective in anticipating the design and usability issues that may result from a real world implementation of MUSEC system. However, although social networks have a high availability and acceptability, learning through them is still in its infancy. To date little is known about the distinct attributes of learning through social mediums in comparison to several traditional learning ways. We intend to carry out extensive research in this direction.